# Transport and trapping of dust particles in a potential well created by inductively coupled diffused plasmas.

Mangilal Choudhary*, S. Mukherjee, and P. Bandyopadhyay

Institute for Plasma Research, Bhat, Gandhinagar – 382 428, India

*mangilal@ipr.res.in

## Abstract

A versatile linear dusty (complex) plasma device is designed to study the transport and dynamical behavior of dust particles in a large volume. Diffused inductively coupled plasma is generated in the background of argon gas. A novel technique is used to introduce the dust particles in the main plasma by striking a secondary direct current (DC) glow discharge. These dust particles are found to get trapped in an electrostatic potential well which is formed due to the combination of the ambipolar electric field caused by diffusive plasma and the field produced by the charged glass wall of the vacuum chamber. According to the requirements, the volume of the dust cloud can be controlled very precisely by tuning the plasma and discharge parameters. The present device can be used to address the underlying physics behind the transport of dust particles, self – excited dust acoustic waves and instabilities. The detailed design of this device, plasma production and characterization, trapping and transport of the dust particle and some of the preliminary experimental results are presented.

## 1. Introduction:

Dusty plasma is an ionized gas that contains dust particles (either dielectric or conducting) with size ranging from tens of nanometers to hundreds of microns. In most of the laboratory plasmas, when these particles are introduced into usual two-component plasma, they acquire negative charge due to the higher mobility of electrons than ions [1]. The small charged dust particles are then suspended near the plasma sheath boundary by balancing mainly the gravitation force and the electrostatic force and form an equilibrium dust cloud. A sheath almost equal to ion Debye length forms around these floating dust particles. These Debye spheres interact with nearby Debye sphere and hence the collection of dust particles has the characteristics of collective nature. The collective properties of this medium provide an identity similar to plasma thus it is termed as "dusty plasma". Such type of plasmas are encountered in industrial plasma [2, 3], space plasma [4, 5], fusion devices [6, 7], and laboratory plasmas [8-24].

In dusty plasma, the charge to mass ratio of dust particles is comparatively smaller than that of the other plasma species. This remarkable property of dusty plasma allows us to study the dynamics of dust particles in the frequency range from few Hz to tens of Hz. Therefore visualization of these particles becomes very easy. Depending upon the amount of charge resides on the dust surface, dusty plasma offers an excellent medium in which one can obtained weakly coupled gaseous state to strongly coupled crystalline state. Therefore, this medium can be used to study the generic fundamental physics of fluid flow, crystalline properties, waves and instabilities in fluid, phase transitions etc. In addition, dusty plasma provides a platform to understand the underlying physics at microscopic levels for various multidisciplinary fields.

However, to understand the physics behind the waves, turbulences, and various instabilities in the dust medium, it is necessary to create dusty plasmas in a length larger than the characteristic length of linear/non-linear modes along with the confinement time longer than characteristics time. It essentially indicates the necessity of a large dusty plasma volume confined for a longer time. Since, the dust particles are negatively charged and hence a positive potential well is required to confine them. Along with this above requirements, a technique or tool is also needed to introduce the dust particles into the electrostatic potential well in a controlled manner. During last couple of decades, various configurations have been made to create an electrostatic trap which enables to confine the dust grains for longer time against gravity.

In some of the experimental dusty plasma devices [8 - 10], which are operated in DC discharge mode, the dust particles are trapped into the anodic plasma which sits nearly 50 – 60 volt above the plasma potential. The dust particles are introduced in the anodic electrostatic trap either using a floating tray or a dust dispenser. The trapped dust particles exhibit 3D dust structure along with associated waves and instabilities. However, these configurations provide the confinement of dust grains but unable to maintain steady state equilibrium below a critical pressure (p < 0 .08 *mbar*), instead various modes appears [8 - 11]. When the pressure increases beyond a threshold value (p > 0.2 *mbar*), the medium shows the ordered structures [12]. Apart from this anodic plasma, standing strata also provides a potential trap to the dust particles in which one can explore the waves and crystallization properties of dust medium [13, 14]. In strata, the dust grains get confined at lower edge and formed a dust cloud of 5 – 20 *mm* in diameter at a higher pressure (p > 0 .1 *mbar*).

In DC discharges, cathode sheath electric field provides steady state equilibrium to levitate the dust grains. This dusty plasma medium is frequently used for studying the Dust Acoustic Waves (DAWs), Dust Transverse Shear Waves (DTSWs), and Dust Acoustic Solitary Waves (DASWs) [15,16]. In this configuration, a steady sate equilibrium dusty plasma volume is only possible when the pressure is more than 0.1 *mbar* otherwise spontaneous DAWs get triggered. Other than these devices, PK-4 [17], a $\pi$ – shaped glass tube that has been employed for studying waves, instabilities and order structures in a dusty plasma at ground level as well as at micro gravity level since past two decades. It is observed in many experiments that equilibrium dusty plasma can be formed in rf discharges even at low pressure (< 0.02 *mbar*). In rf configurations, dust particles are confined at sheath edge of powered electrode with an additional confinement ring installed at lower electrode. Since, sheath electric field decreases monotonically from the electrode surface to the plasma which allows dust grains to form a 2D dust layer above the rf electrode [18, 19]. However, 3D dust structure with less extent (nearly 10 *mm*) is formed above the rf electrode which still suits to perform the wave experiments [20, 21]. A larger 2D dusty plasma crystal can be formed in rf operated discharges at lower pressure for wave and flow studies [22, 23]. Apart from these above methods, diffused edge of inductively glow plasma is used to confine the dust particles by Fortov et al. [24].

According to earlier experimental observations, an equilibrium dusty plasma can be created at lower pressure in rf discharge plasma but the gravity leads to particles sedimentation into a 2D flat dust cloud. It is also observed that most of the dusty plasma studies have been carried out in uniform and non-uniform plasma background inside a small 3D dust cloud. Still an extensive study on self – excited turbulences, dust – acoustic instabilities, driven nonlinear modes, dust dynamics in non-uniformed dust column, transient response and dynamical effect on dust particles during relaxation process in a 3D dust cloud at lower pressure is unexplored. Along with a stable confinement, it is also necessary to introduce the dust particles in the experimental region in absence of ion and neutral drag forces so that the volume of dust particles can be tuned precisely during the experiments.

Keeping all these issues in mind, we have built a simple device capable of producing large 3D dusty plasma of maximum dimension of $35 \times 220$ *mm$^2$* in Y− Z and $30 \times 220$ *mm$^2$* in X – Z plane in the non-uniform diffused plasma at lower pressures. Also, a novel technique is followed to introduce the dust grains into the potential well of the main plasma volume with the help of ambipolar electric field. It also provides the flexibility on the control of dust density and inter-grain distances. Details of experimental device and techniques are mentioned in Section 2 of this article. Section 3 describes the plasma diagnostics and plasma potential variation in diffused plasma. Dust transport and trapping is described in Section 4. Conclusion is summarized in the Section 5.

## 2. Experimental device and techniques:

The cylindrical linear dusty plasma experimental device (shown in Fig. 1(a)) is made of borosilicate glass tube of length 60 *cm* and of diameter 15 *cm*, has six radial and two axial ports for different purposes. Gas injection and pumping systems are attached with the experimental chamber through a Stainless Steel (SS) tube (buffer chamber), which is of 20 *cm* in length and 15 *cm* in diameter. This assembly prevents the direct gas flow in the main experimental chamber where the dusty plasma experiments are carried out. Therefore, the contribution of flow induce neutral drag on the dust particles is neglected. Schematic of the experimental assembly is shown in Fig. 1(b).

To begin with we define the coordinate system of our experimental setup. In this experimental configuration Z = 0 cm and Z = 60 cm correspond to the left and right axial ports (as shown in Fig. 1), respectively. X = 0 cm and Y = 0 cm indicate the points on the axis passes through the center of the tube. The center of source tube is located at Z ~ 12 cm whereas the dust reservoir is located at Z ~ 45 cm.

A SS disk of 6 *cm* diameter is used as a dust reservoir and mounted at one of the radial ports (Z = 45 *cm*) of the chamber by a cylindrical rod. The disk has a step like structure of 5 *mm* width and 2 *mm* height to confine the particles. There is a provision to change the radial position of the dust reservoir. Before closing the chamber, the dust particles are sprinkled on the disk homogeneously. The experimental chamber is evacuated at ~ $10^{-3}$ *mbar* pressure using a rotary pump and subsequently the argon gas is fed into the chamber. Then the chamber is pumped down again to the base pressure. This process is repeated three times to reduce the impurities from the vacuum chamber. Finally the operating pressure is set to 0.03 − 0.1 *mbar* by adjusting the gas-feeding valve.

A copper made spiral ring (of 5 turns) is used as RF antenna, which is wound on the source tube (12 cm long and 8 cm diameter) connected with the experimental chamber as shown in Fig.1. An inductively coupled radio frequency (of 13.56 MHz) discharge is initiated inside the source tube by coupling a RF power (ranging from 2 – 7 watt) through a matching network. It is observed that the plasma enters into the experimental chamber from the source tube and then diffuses along the axis of the chamber. The typical views of diffused plasma glow in Y – Z and X – Y planes are shown in Fig. 2(a) and 2(b). This glow boundary changes with the change of discharge parameters.

To inject the kaolin dust particles ($\rho$ ~ 2.6 *gm/cm³* and $r_d$ ~ 0.5 to 5 *μm*) into this plasma, the dust-containing disk is biased negatively (about -300 V or above) to form a dc glow discharge around the disk.

Dust particles get charged, lifted up and trapped at the cathode sheath edge. Even though the particles are trapped but because of the poor confinement and/or high voltage cathode spots, they leave the trapped region and spread into the plasma volume. As these charged particles reached to the edge of diffused plasma, they drifted and confined in to the electrostatic potential well formed by the inductively diffused plasma. It is worth mentioning that the plasma glow intensity (see Fig. 2(a) and Fig. 2(b)) in presence of dust grains reduces slightly, as seen by the eye, probably due to the electron absorption on the dust surface. However, this change cannot be seen properly in the camera images. The mechanism of dust transport and trapping will be discussed in more details in subsequent sections.

These dust particles are illuminated by the combination of a tunable red diode laser (632 nm wavelengths, 1−100 mW power and 3 mm beam diameter) and cylindrical lens (plano−convex). The dynamics of the dust particles are then captured by a Lumenera make CCD camera having frame rate of 130 fps and spatial resolution of 1088 × 2048 pixels. A standard Zoom lens of variable focal length (from 18 *mm* to 108 *mm*) is also used for the magnification purpose during the experiments. The maximum field of view at minimum zoom mode is ~ 150 × 75 $mm^2$ and it reduces to ~ 25 × 12.5 $mm^2$ at maximum zoom mode. There is a provision to image the dust cloud either in horizontal (X−Y) or in vertical (Y−Z) planes by changing the orientation of the cylindrical lens and camera. The series of images are then stored into a high-speed computer and later analyzed with the help of Matlab based freely available PIV software.

## 3. Characterization of Inductive diffused plasma:

The diffused plasma in the main experimental chamber is characterized thoroughly by different electrostatic probes namely, the single Langmuir probe [25-28], double probe [29] and emissive probe [30]. A rf–compensated cylindrical Langmuir probe of length 7 *mm* and radius 0.25 *mm* is used to measure the plasma density ($n_e$) and the electron temperature ($T_e$) in a wide range of input rf powers (3 – 7 Watt) and neutral gas pressures (0.03 − 0.2 *mbar*). Fig. 3 shows the variation of plasma density and electron temperature in axial (see Fig. 3(a)) as well as in vertical (see Fig. 3(b)) directions at P = 5 W and p = 0.05 *mbar*. In the axial profile (at Y ~ 0 *cm*) as shown in Fig. 3(a), the electron density initially increases from ~ 4×$10^8$ $cm^{-3}$ (at Z ~ 1 *cm*), takes a maximum value of ~ 7×$10^8$ $cm^{-3}$ (at Z ~ 12 cm) and then finally falls monotonically to ~ 1×$10^8$ $cm^{-3}$ (at Z ~ 30cm). The axial electron temperature profile shows a similar trend like plasma density. The temperature in axial direction changes from a maximum value of ~ *8* eV to a minimum value ~ 3 eV. The variation of $n_e$ and $T_e$ in vertical direction (at Z ~ 12 cm) is shown in Fig. 3(b). It is seen the maximum values of density (~7×$10^8$ $cm^{-3}$) and temperature (~8 eV) is measured approximately at the center of the experimental chamber. Both the density and temperature then decrease from these peak values when the probe is scanned toward the wall of the experimental chamber.

To study particle trapping and transport in the background of inductively coupled diffused plasma in more detail, it is essential to study the potential profiles structure. To begin with, the floating and plasma potentials are measured using an emissive probe in a wide range of discharge parameters at different locations. The simplest method, i.e., *floating potential* method (see ref. [30] and references therein), is followed to measure these potentials. The cold emissive probe acts as a single Langmuir probe and gives the floating potential ($V_f$) at zero emission current. The floating potential shifts towards the plasma potential ($V_p$) when the probe starts emitting the electrons. With the increase of emission current, the floating potential rises rapidly and then reaches a constant value. This value is known as plasma potential, which does not change with further increase of emission current.

Fig. 4 displays the variation of plasma potential along the axis of the experimental chamber for a given location, X = 0 *cm* and Y = - 3 *cm*, at different disk bias voltages. For these measurements the background neutral gas pressure ($p$ = 0.05 *mbar*) and electrode power ($P$ = 5 Watt) are kept constant. It is clearly seen in Fig. 4 that the measured plasma potential decreases on both the sides if one goes away from the potential well localized at Z ~ 12 *cm*. The change in plasma potential becomes insignificant beyond Z ~ 30 *cm*. This measurement implies that there is a strong gradient in plasma density, which essentially creates a potential well centering at Z ~ 12 *cm*. It is also to be noted that the depth of the potential well remains almost constant whereas the magnitude of the plasma potential decreases with the increase of the applied bias voltage to the disk.

Fig. 5(a) shows the plasma potential variation along Z – axis (similar to Fig. 4) for two different neutral gas pressures $p$ = 0.05 *mbar* (indicated by squares) and 0.025 *mbar* (indicated by closed circles) at a given rf power of 5 watt whereas Fig. 5(b) shows the same for two different rf powers ($P$ = 5 and 6.5 watt) at a particular pressure, $p$ = 0.025 *mbar*. In both the cases the DC plasma is kept switched off. It is noticed from the Fig. 5(a) that the depth of the potential well increases whereas the width decreases significantly with the increase of the gas pressure. Interestingly, the center of the potential well also shifted from left (Z ~ 11 *cm*) to right (Z ~ 16 *cm*) when the pressure is decreased from 0.05 *mbar* 0.025 *mbar*. However, the depth and the width of the potential well increase when the electrode power is increased from 5 Watt to 6.5 Watt keeping the center at the same location.

As discussed in Sec. II, the negatively charged dust particles get levitated near the region where the upward electrostatic force caused by the diffuse plasma is exactly balanced by the gravity. To study the potential profiles along the direction of the gravity, the plasma potential is plotted along Y-direction for different values of X location as shown in Fig 6. The different X locations are X = -3, 0, and 3 *cm*, respectively for a given Z ∼ 12 *cm*. The figure is plotted for the discharge parameter, *p* = 0.05 *mbar* and *P* = 5 Watt. For this above discharge condition, it is clearly seen in Fig. 6 that the plasma potential follows a nearly symmetrical trend in vertical direction from center for a given X location. The plasma potential decreases gradually in the direction of gravity (along and opposite) if the probe is scanned away from the center. It is also to be noted that the plasma potential remains almost constant near the center of the experimental tube (for a particular X and Z location). It is also worth mentioning that the plasma potential decreases when the probe is scanned from X = - 3 (close to the plasma source) to +3 (away from the source) *cm* for given Z and Y values due to the gradient of plasma density along X−axis. The estimated error in the measured value of plasma potential is within the range of 1-2 ($\pm$ 5 %) volts.

## 4. Dust grain transport and trapping:

In most of the experimental setups [15 - 22], the dust particles get negatively charged by collecting more electrons than ions and levitate at the cathode (in case of capacitive coupled rf, it is either grounded or powered electrode) sheath boundary by balancing the gravitational and electrostatic forces. In this device, a unique technique is followed to introduce charged dust particles in the diffused plasma. The dust reservoir is biased negatively (about -300 V or above) to produce a secondary DC glow discharge plasma. Acquiring more electrons than ions from the background plasma the dust particles become negatively charged. These negatively charged dust particles then levitate near the sheath boundary by balancing electrostatic force and the gravitational force similar to earlier experiments [15-22]. These weakly confined dust particles are then transferred into the main plasma volume by adjusting the discharge parameters of the DC glow. When the particles fall down at the edge of the potential well shown in Fig. 4, they drift towards the center of the well as discussed in earlier section. After entering into the main plasma, these dust particles transport towards the source chamber and finally get confined at a particular axial location where the confinement is strong enough. The novelty of our experimental device is to levitate these negatively charged grains in the electrode less sheath in the experimental regime. The vertical component of the electric field, analogous to the sheath electric field comes from the diffused plasma, helps us to lift the dust particles against the gravity. Sometimes the micro arcs at higher pressure/electrode voltage in the sheath (mainly in the cathode sheath) region disturb the plasma and hence the dynamics of the dust particles get affected. Essentially it restricts us to perform experiments in these specific discharge conditions near the cathode sheath, which can be avoided easily in this device.

After the confinement of the particles in the main plasma, the DC glow discharge plasma is switched off to provide even a better confinement to the particles. It is because the dust particles are transported toward central region while DC plasma is on. The axial confinement potential in absence of the DC glow discharge plasma is shown in Fig. 5 for two different pressures and rf powers. To estimate the electric field (E) to levitate the dust particles of diameter ranging from 1 to 10 $\mu m$, the force (the gravitational force to the electrostatic force) balance condition [31] can be used as follows:

$$E = \frac{r_d^2 \rho g}{3\varepsilon_0 V_s} \tag{1}$$

Where, $r_d$ is radius of the dust particle, $\rho$ is mass density of kaolin, g is gravitational constant, $\varepsilon_0$ is dielectric constant of vacuum and $V_s$ is dust surface potential.

For kaolin dust particles ($\rho \sim 2.6\ gm/cm^3$ and $r_d \sim 0.5$ to 5 $\mu m$), the required electric field turns out to be ~ 1V/cm to 12 V/cm. Based on the plasma potential gradient, shown in Fig. 6, the calculated electric field varies from 0.5 to 4 V/cm at different heights and radial locations. Hence, the dust particles (kaolin particles with average radius of 3 $\mu m$ or MF particles with average radius of 4 $\mu m$) can be levitated at different X and Y locations for a particular Z location depending upon the force balance condition.

The velocities of the flowing particles can be estimated by using Particle Image Velocimetry (PIV) analysis. In this technique, the velocity profile can be constructed by performing cross – correlation between two consecutive frames [32, 33]. The images with illuminated dust grains are decomposed into a similar interrogation boxes with dimension 64 × 32 pixels. After that a cross – correlation of two images of a single interrogation box is used to construct the velocity vector for dust particles lying in the interrogation box. The distance travel by a particle in the two consecutive frames is known therefore velocity of flowing dust particles can be computed. For the measurement of drift velocity of the particles, we have performed the experiments at the background gas pressure, $p = 0.05\ mbar$ and rf power, $P = 5$ Watt and considered few consecutive frames of flowing dust particles.

Fig. 7(a) shows a typical image of PIV analysis, where the direction of the arrows shows the direction of the flow whereas the length of the arrow shows the magnitude of the velocity. It is clear from Fig. 7(a) that the particles are drifting towards the center of the well not horizontally rather obliquely. Although, the vertical component of the velocity is negligible compared to the horizontal component. Fig. 7(b) shows quantitative variation of these velocity components, $u_z$ and $u_y$, along the Z and Y−axes. It is noticed from this above figure, the velocity (average value is 0.23 $cm/sec$) in Y − direction remains almost constant whereas the velocity varies from 1.6 cm/sec to 1.0 *cm/sec* along the axis of the experimental tube. The possible explanation is provided in the subsequent section.

When we switch off the applied DC voltage to the dust reservoir, the transport of the dust particles stops and they get confined in a region where they satisfy the force balance condition as discussed above. As a result a stable long 3D dust cloud forms. A 2D slice of dust cloud (3D) in vertical plane (Y – Z) is presented in Fig. 8. It is observed that the dust cloud expands at lower pressure and the shifting of its center toward the reservoir if the other parameters are kept constant. The dust cloud expansion and its center shifting can be explained on the basis of broadening of the potential well at lower pressure (Fig. 5(a)). It is also observed that the expanded dust cloud can be achieved with increasing the input power at given pressure which is directly related to the broadening of potential well (Fig. 5(b)). Hence, it is concluded that by adjusting the gas pressure, input rf power, and DC bias to the dust reservoir, dust density as well as the dust cloud volume can be controlled.

To study the dynamics of the dust particles, we have estimated the possible forces acting on the particle during the transport and the confinement. For this study, we perform an experiment with kaolin particles of average radius, $r_d \sim 3 \mu m$ and mass density $\rho_m \sim 2.6 \, gm/cm^3$. The gravitational force experienced by the particle comes out to be $F_g \approx 10^{-12} N$ for the particles of average mass $m_d = 2.8 \times 10^{-13}$ *kg*. As discussed, during the confinement, the downward gravitational force is mainly balanced by the upward electrostatic force. For a spherical dust sphere, the charge acquired on the surface of the dust can be expressed as:

$$Q_d = 4\pi\varepsilon_0 r_d V_s \qquad (2)$$

where, $V_s \sim -4T_e$ (V) is the dust surface potential which can be estimated from the Orbital Motion Limited (OML) theory [1] at net zero current limit on the dust surface. $T_e$ is electron temperature in the electron volt unit. For our experimental conditions, the dust charge number ($Z_d = \frac{Q_d}{e}$) comes out to be $Z_d \approx 3 - 8 \times 10^4$. However, it is to be noted that, this simplest approach using OML theory always overestimates the dust charge [34] in laboratory condition. The electrostatic force for the measured average electric field E = 3*V/cm* is $F_E = Q_d E \approx 10^{-12} N$. It implies that the gravitational force is well balanced by the electrostatic force for the present experimental condition to levitate the micro particles. We have estimated other responsible forces that can act on the particles during the levitation. The ion drag force acting on the charged dust particles is estimated by Barnes formula [35], which comes out to be $F_i \approx 10^{-13} N$. This force is almost an order less than the gravitational and the electrostatic forces. Although, the OML theory sometimes underestimates the ion drag force as predicted by Khrapak et al. [36]. In spite of, it is used to determine the charge and the ion drag force for the sake of simplicity. Hence we can conclude that the ion drag force has insignificant role to levitate the particles inside the potential well. The rf-coils on the source tube can heat up the glass chamber even at lower power (2 – 7 W). As a result, the temperature gradient induced thermophoresis force [37, 38] can affect the equilibrium of dust cloud. However, our measurements are made far from the source tube where the contribution of thermophoresis force on the dust particle is considerably smaller compared to other dominating forces. Hence, we ignore the effect of thermophoresis in our calculations.

As discussed earlier that the particles can be transported from DC plasma source to diffuse plasma by adjusting the biased voltage applied to the dust reservoir. During the transportation, the dust particles flow with a velocity ranging from 0.5 to 3 *cm/sec* depending upon the discharge condition and the axial location. This transport of the particles can be explained by the combination of the forces that are acting on the particle. The main responsible forces experienced by the particles during the flow are the axial electrostatic force, ion drag force, neutral drag force and the Coulomb repulsive force. The electrostatic force on the dust particles arises due to the gradient of axial plasma potential, which drags the particle towards the center of the potential well. Whereas the other three remaining forces always restrict the motion of dust particles. The ion drag force, appears because of the gradient in the plasma potential, tries to drag the particle along its own direction. The flowing neutrals can also carry the dust particles along its direction. In absence of neutral flow, the stationary background gas provides the neutral drag force which always resists the motion of the particles. The Columbic opposing force between the drifted particles and the stationary dust cloud plays also an important role to determine the dust dynamics during the transportation. It is clear from Fig.7, that the drift velocity decreases gradually mainly because of this repulsive force that comes from the confined dust particles. It is also observed that the drifted dust particles become almost stationary due to this strong repulsive force near the center.

A series of experiments can be performed in this tabletop versatile experimental chamber apart from the transport in the diffused field. A 3D large equilibrium dust cloud of 35 × 130 $mm^2$ in Y − Z and 30 × 130 $mm^2$ in X – Z plane is observed when the experiments are carried out at low pressure (0.04 *mbar*) and low power (4W). Fig. 8 shows Y−Z plane of this 3D equilibrium dust cloud. It is also observed in our experiments that the volume and the location of the dust cloud can be changed by changing the discharge parameters. Since the potential well created in this configuration provides a better confinement to the dust particles therefore we can perform experiments for a longer time with a constant dust density. The dynamical studies of such equilibrium dust grains with external driving force will be addressed in the upcoming publications.

The stable dust cloud shows the propagation of spontaneous dust acoustic wave in the direction of the gravity when the dust density and/or the rf power is increased from the equilibrium condition. Fig. 9 shows a typical view of spontaneous dust acoustic waves seen in our experiments. Sometimes it is also observed that the linear wave gets bifurcated and/or merged during the propagation along its direction. Encircle regions in Fig. 9 represent the wave front splitting and merging which may be the results of defect mediated instability as reported in ref. [21]. The future study of this observation will be reported in forthcoming publications.

## 5. Conclusions:

In this paper, a newly built versatile experimental device is presented for studying different phenomena like waves, instabilities, and dynamical behavior of dust particles with transient perturbation in low pressure dusty plasma. A secondary dc plasma source is used to inject the dust particles into inductively coupled diffused plasma which is a unique feature of this device. The axial component of the ambipolar electric field transports the injected dust particle into an electrostatic trap where a 3D dusty structure is formed. Another aspect of this device is the non-uniformity of plasma density in radial and axial direction, which essentially leads to the formation of a non-uniform long dust column that allow us to perform some specific experiments on the wave propagation. Additionally the transport of charged particles can also be used to explore the studies on transport phenomena of medium and flow profiles around an additional boundary in the path of flow. This electrode less dusty plasma experimental device is free from micro arcing, which can sometimes affect the dynamics of dust particles. Initial experiments to characterize the diffused plasma, dust transport and trapping mechanism, and some preliminary results on wave modes have been reported.

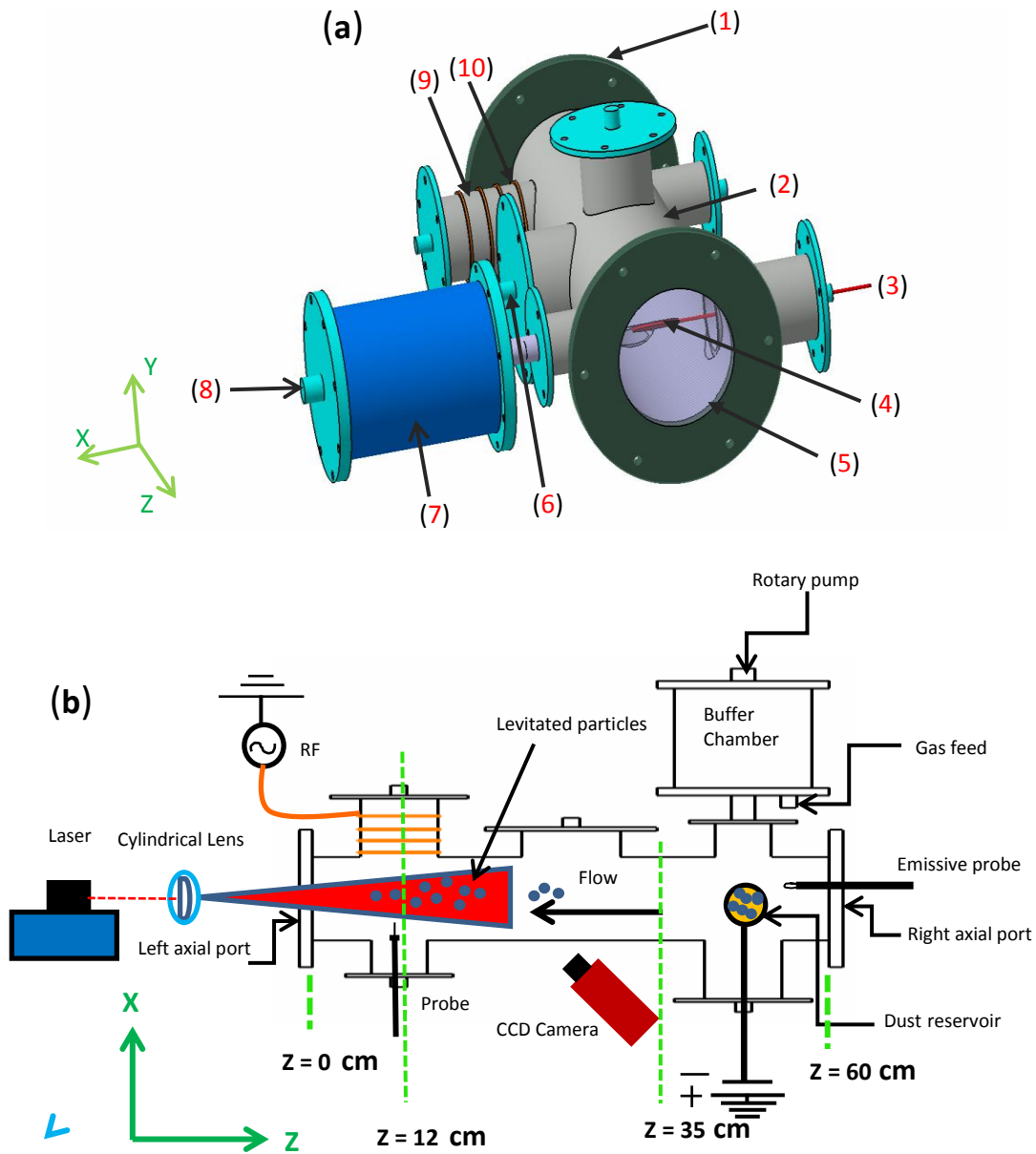

FIG.1 (Color online) (**a**) 3D view of experimental assembly: (1) Left axial port. (2) Experimental chamber. (3) SS rod for holding the dust containing disk. (4) Dust containing disk (dust reservoir). (5) Right axial port. (6) Gas feeding valve (attached to buffer chamber). (7) Buffer chamber. (8) Rotary pump. (9) ICP plasma source tube. (10) Inductive coils. (**b**) Schematic of Inductively coupled dusty plasma system (Top view).

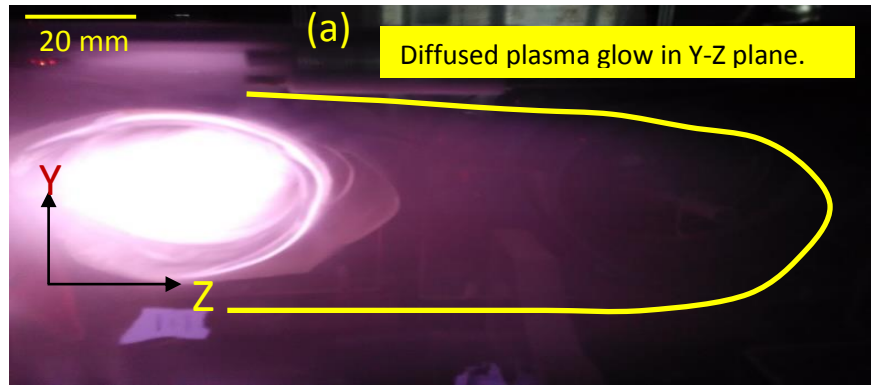

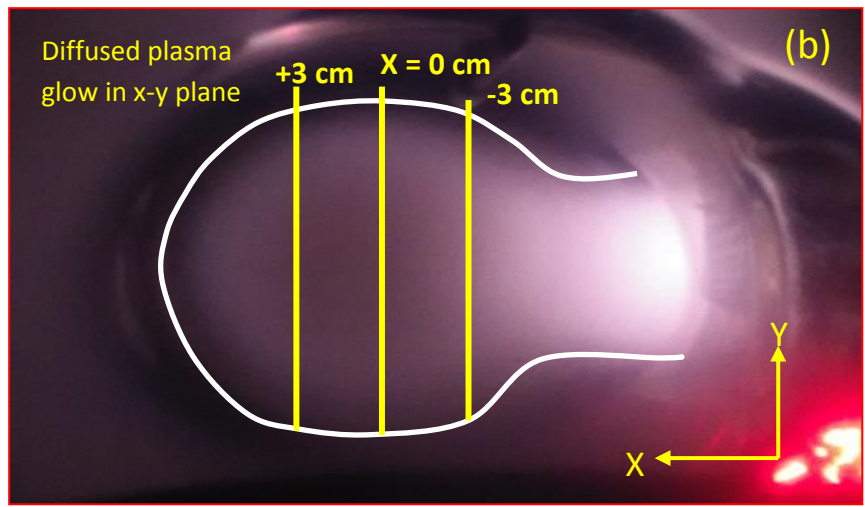

FIG.2 (Color online). (a) A typical plasma glow in Y−Z plane. Yellow line represents the boundary of glow region. (b) Plasma Glow profile in X −Y plane. White line indicates the boundary of glow region in this plane. The vertical yellow lines represent the different vertical positions for given X – value. Argon pressure and rf power are 0.05 *mbar* and 5 Watt, respectively.

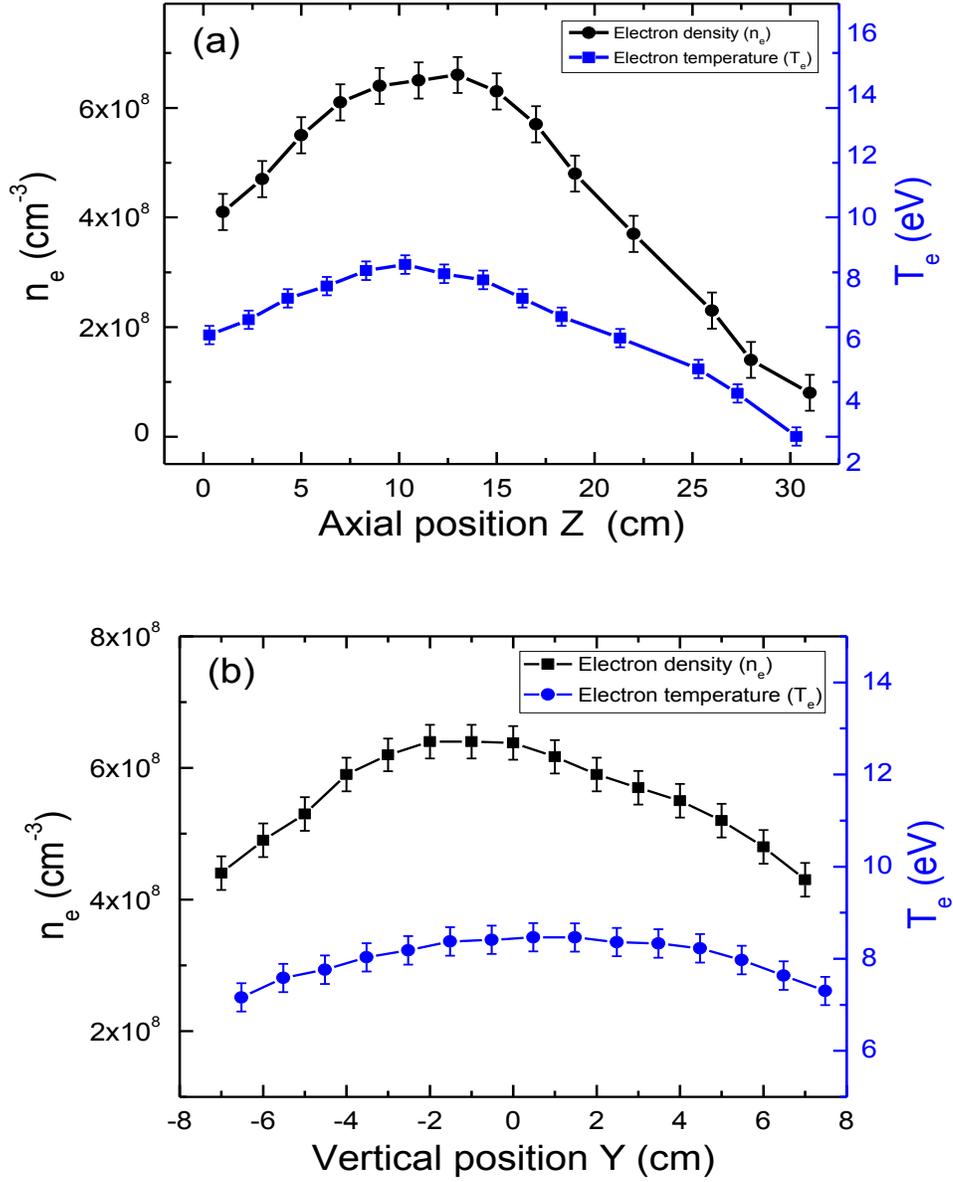

FIG. 3 (Color online) Electron density ($n_e$) and temperature ($T_e$) profiles: (a) along Z - axis of the tube (X ~ 0 and Y ~ 0 cm) (b) along Y – axis (X ~ 0 and Z ~12 cm). Argon pressure and rf power are 0.05 mbar and 5 W, respectively. The errors in density and temperature measurements are within $\pm 3 \times 10^7$ cm$^{-3}$ and $\pm 0.3$ eV, respectively.

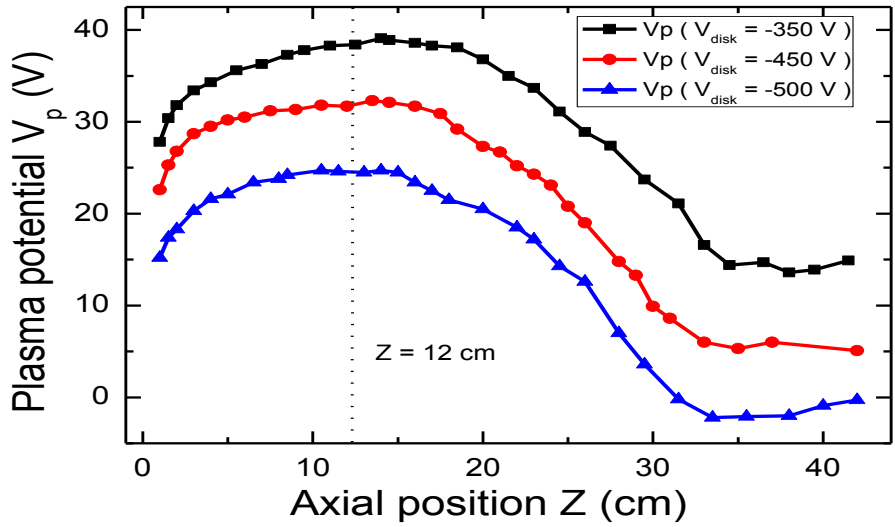

FIG. 4 (Color online) Plasma potential profiles along the axis of the tube for different disk bias voltage at neutral pressure, $p = 0.05$ mbar and rf power, $P = 5$ W. Dotted line represents the central region where dust particles get confined during the transport. The measurement errors are within ± 5%.

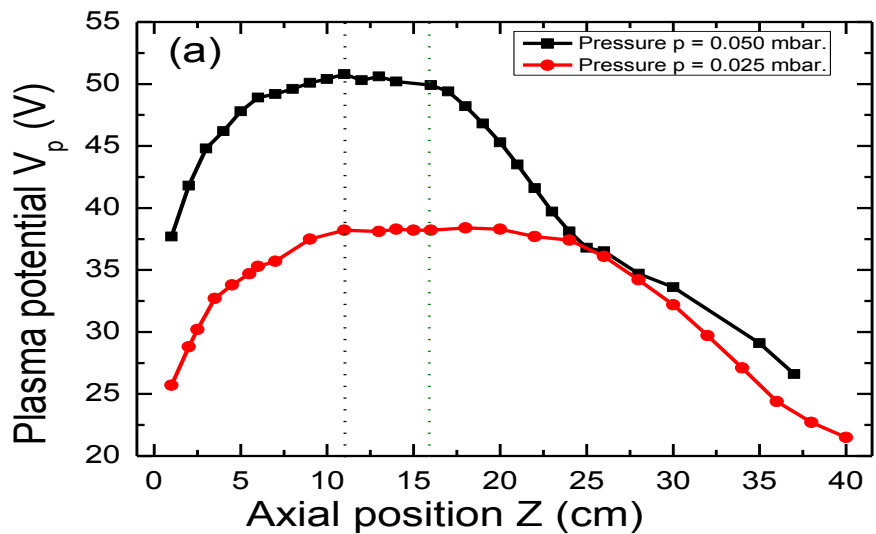

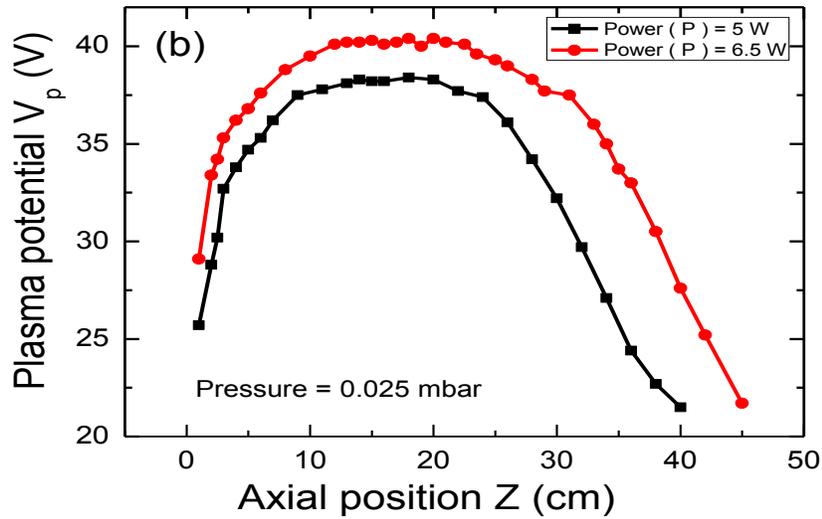

FIG.5. (Color online) Axial plasma potential profiles of diffused plasma for (a) two different gas pressures $p = 0.05$ and $0.025$ mbar at a fixed rf power $P = 5$ $W$ $and$ (b) two different rf powers P = 5 and 6.5 watt at a fixed gas pressures, $p = 0.025$ mbar. The measurements are taken after switching off the DC plasma. Black and green dotted lines (in Fig. 5 (a)) represent the centre of the potential well at 0.05 and 0.025 mbar, respectively. The measurement errors are within ± 5%.

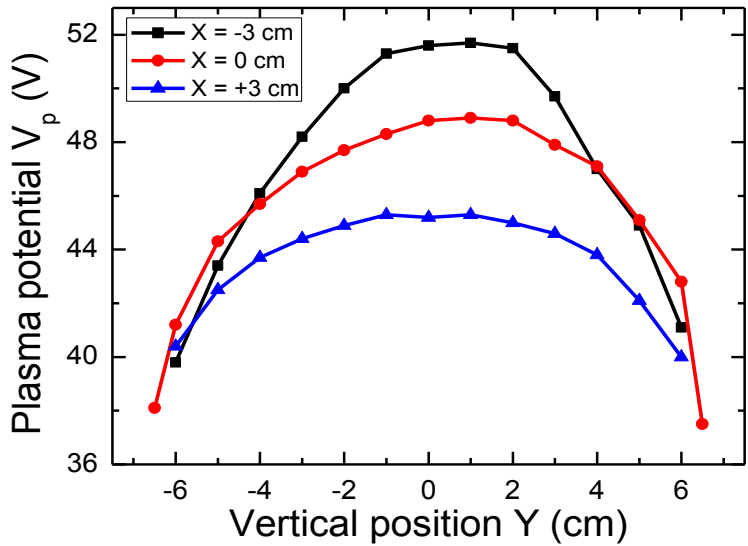

FIG. 6 (Color online) Plasma potential variations in different vertical planes of a radial cross section near the central region (in X – Y plane). Potential traces are corresponds to different X – values. Argon gas pressure and rf power are 0.05 mbar and 5 W, respectively. The measurement errors are within ± 5%.

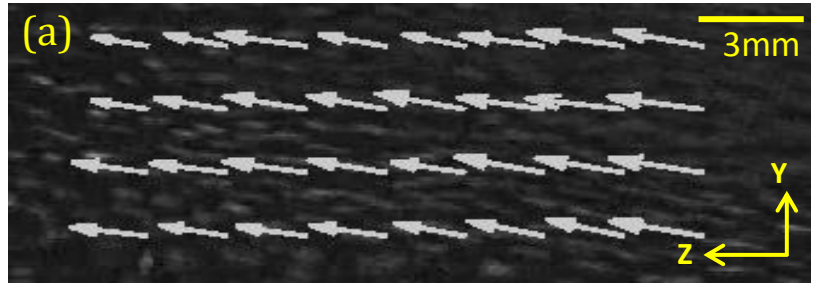

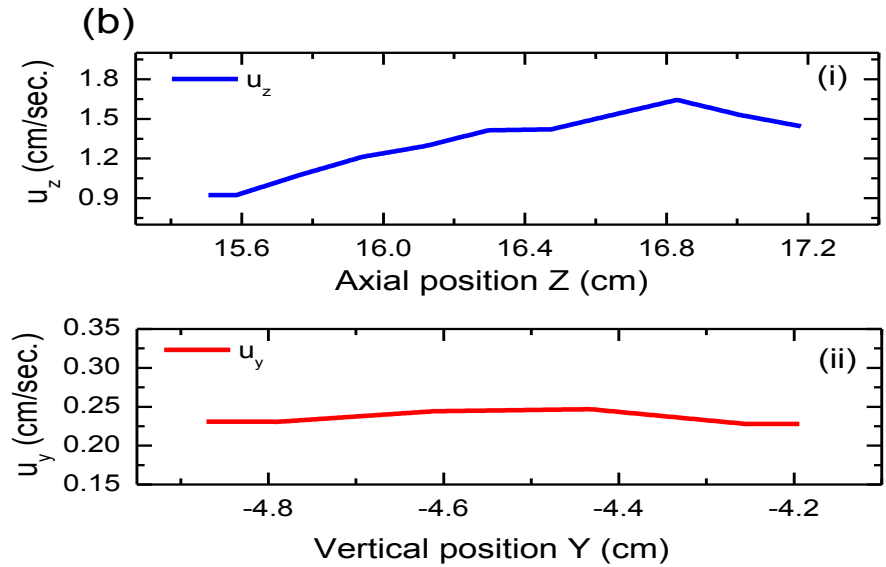

FIG. 7 (color online) (a) PIV analysis of kaolin particles which shows the flow of particles toward the central region (Z = 12 cm) when the dust reservoir is biased to -450 V. The direction of arrows indicates the dust particles flow whereas length corresponds to the magnitude of velocity vector. The white spots in the image correspond to the flowing kaolin particles. (**b**) Velocity profiles in two directions (Z and Y – directions). During the flow argon pressure and input power are 0.05 mbar and 5 watt, respectively.

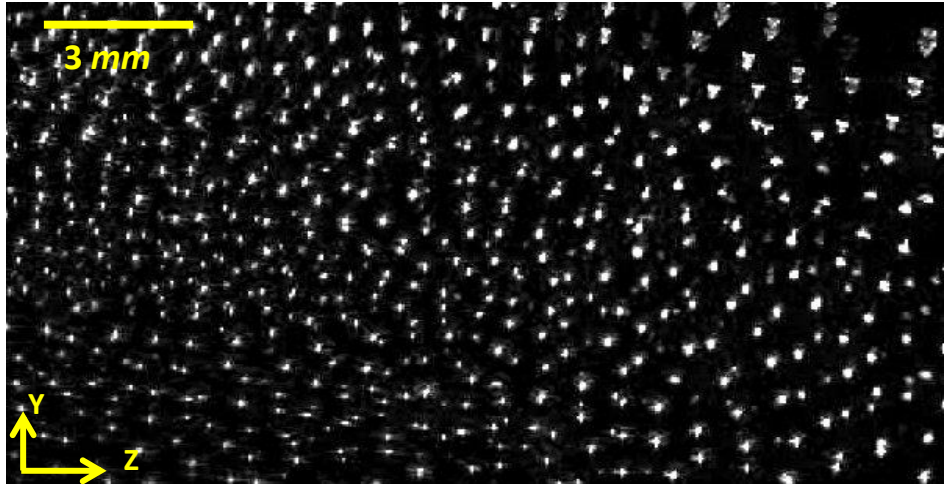

FIG.8 (Color online) A snapshot of dust particles distribution in Y – Z plane (Z ~ 13 cm). Bright points correspond to the dust particles. Kaolin particles with average radius of 3 *μm* are used in this experiment. Argon gas pressure is set to 0.04 mbar and input rf power is 4 W. The field of view is $18.64 \times 11.5\,mm^2$.

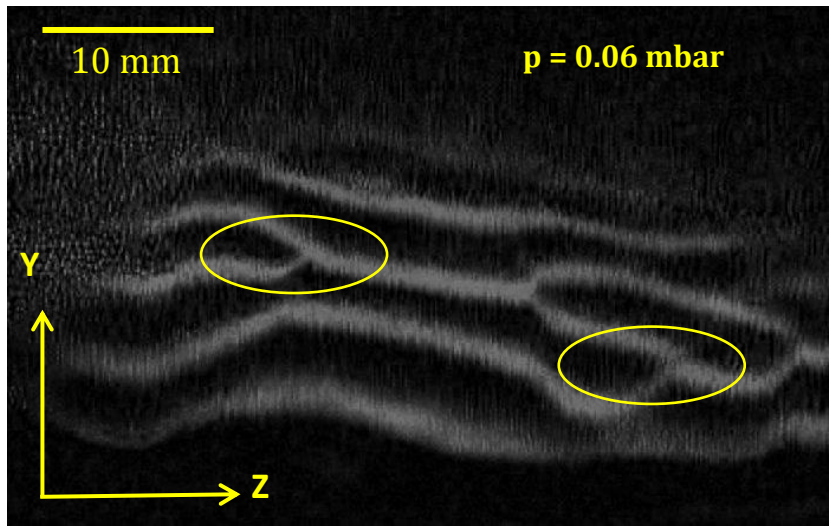

FIG.9 (color online) A single video frame of dust cloud in Y – Z plane near the central region shows the breaking and merging of dust acoustic wave at pressure $p = 0.06$ *mbar* and rf power P = 5 watt. For this experiment kaolin particles of average size ($r_d$ ~ 3 *μm)* are used. The field of view is $46.8 \times 25.6\,mm^2$.